\documentclass[preprint,aps,onecolumn,showpacs,prb,12pt]{revtex4}
\usepackage{graphicx}
\usepackage{amsmath}
\usepackage{natbib}
\usepackage{epsfig}
\usepackage{amsfonts}
\usepackage{amssymb}

\begin{document}

\title{Spin currents in the Rashba model in the 
presence of non-uniform fields}

\author{A. Rebei}
\affiliation{  Seagate Research Center, Pittsburgh, PA 15222}

\author{ O. Heinonen}
\affiliation{  Seagate Technology, Bloomington, MN 55435}


\begin{abstract}
Spin currents in a two dimensional electron gas with Rashba-type 
spin orbit
coupling are derived from a spin connection. \ Using a functional 
integral
method, we recover the result derived by 
Sinova {\em et al.}$\left[ \mathrm{Phys.Rev. Lett.  
 92, 126603 (2004)}\right]$
for a uniform 
electric field and in the absence of impurities. \ We extend this 
result  to
inhomogeneous electric and magnetic fields. \ We find that 
non-uniform 
magnetic fields can  give rise to spin currents that 
are independent 
of the Rashba coupling and hence are less susceptible to 
impurities than 
in the case of uniform electric fields.
\end{abstract}
\maketitle
\begin{figure}
\mbox{\epsfig{file=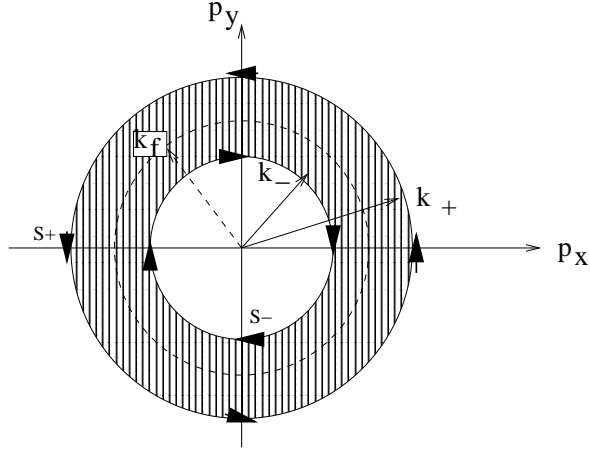,height=6 cm}}
\caption{Fermi surfaces of the two bands in the Rashba model, $k_{\pm}= k_F \pm m \lambda$. \ The tangential arrows are the spin projections
 on the $k_xk_y$ plane. \ The point $k=0$ is a singular point.}
\label{FermiSurface}
\end{figure}


\ Spin-orbit coupling (SOC) of the conduction electrons 
in two dimensional
systems is of special importance in magnetic and 
semiconductor materials.  \ It was recently realized
that SOC can be used to manipulate the spin of the conduction 
electrons in semiconductors. \ An
interesting outcome of this coupling is that an equilibrium 
spin current in
semiconductors becomes possible due to SOC in the presence of 
an in-plane
electric field \cite{das}. \ This topic is
currently of considerable interest since the question on the 
size of this
current is still open. \ One of the reasons for this lack of 
agreement is that
there is no rigorous definition for these currents because they are not
conserved in a medium with a spin orbit coupling \cite{rashba1,rashba2}.

\ In this note, we point out  a well defined procedure for 
calculating spin
currents through first defining a spin connection \cite{FS} 
similar to the
procedure of defining charge currents with respect to a $U(1)$ 
connection.
\ We discuss the simple two dimensional system of an electron gas with
spin-orbit Rashba-type coupling \cite{sinova}. \ The Rashba coupling 
describes
well the dynamics of conduction electrons in semiconductors, e.g., GaAs, which
are potentially important materials for spintronics.\ Similarly, 
the Rashba
model is of interest to conduction electrons in magnetic thin 
films and the
damping problem of the magnetization in these materials. \ We 
will 
treat two cases; first a 2-dimensional electron gas with SOC 
in the presence of an electric field directed in-plane, 
and second with an inhomogeneous magnetic field and electric field. 
  \ The first case 
has already been treated by Sinova {\em et al.}~\cite{sinova} and 
others, while
the non-uniform magnetic field case is new. \ It also has
a universal behavior and hence less susceptible to be 
destroyed by impurity scaatering.

\ The recent result by Sinova {\em et al.}~\cite{sinova} that  
spin
currents are possible in systems described by Rashba coupling 
attracted considerable
attention \cite{inoue,raimondi,loss,chalaev,bernevig,chao,chao2,niu,sun}
because of its universal features  . \ The value of $|e|/8 \pi$ for
 the spin Hall
conductivity is independent of the strength of the SOC. \ Since spin is not 
conserved in this
system, much debate has been centered around the question 
of what is the
correct way to define the corresponding spin current. \ In the 
following, we argue
that the best way to approach the question of what is the 
appropriate equation
for the spin current, is by first identifying the associated 
$SU(2)$ spin
connection to the current, if such one
 exists. \ Then, we calculate 
the effective
action of the 2DEG from which a definition of the spin 
current follows
unambiguously by differentiation with respect to the $SU(2)$ 
connection. \ This
procedure is well known in Gauge theory \cite{schwinger,IZ}.

\ The Rashba Hamiltonian for a two-dimensional electron gas 
(2DEG) is given by
\cite{bychkov}
\begin{equation}
H_{0}=\frac{1}{2m}\left(  p_{x}^{2}+p_{y}^{2}\right)  +\lambda\left(
p_{y}\sigma_{x}-p_{x}\sigma_{y}\right) ,
\end{equation}
where $\lambda$ is a coupling constant, $m$ is the effective 
mass of the
electron in the lattice, $\mathbf{\sigma}$ is a Pauli matrix
vector, and we
use units such that $\hbar= c = 1$. \ The spin vector of the 
electron is
$\mathbf{S}=\frac{1}{2}\mathbf{\sigma} $ and its magnetic moment is
$\mu=-\frac{e}{2 m} \; (e < 0)$. \ The action of this system is 
clearly $U(1)$
invariant and this gives a continuity equation for the 
charge. \ Therefore, we
naturally seek a continuity equation for the spin by first 
finding under what
conditions this system has a $SU(2)$ symmetry which is 
generated by a spin
charge. \ This question was first asked by Frohlich and 
Studer \cite{FS} for a
general non-relativistic particle interacting in an 
electromagnetic field.
\ They were able to show that such systems exhibit a $U(1)\times SU(2) $
symmetry and hence a corresponding charge current and spin current follow
directly from this gauge symmetry. \ Following this procedure, we can ask
similar questions for the Rashba system and this will allow us to identify a
spin connection and the corresponding covariant derivatives. \ It is well
known from Gauge theories, that currents with sources will not be continuous
but only \textit{covariantly} continuous.

\ In mathematical terms, we want to see if it is possible to express the
equation of motion for the electrons in the following form
\begin{equation}
i\frac{\partial\psi}{\partial t}=-\frac{1}{2m}\left(  \mathcal{D}%
_{k}\mathcal{D}^{k}\right)  \psi+e A_{0}\psi+ g \mu\mathbf{V}_{0}%
\cdot\mathbf{S}\psi,
\end{equation}
where $g=2$ for an electron in vacuum and 
the spatial covariant derivative is $
\mathcal{D}_{k}=\nabla_{k}+i e A_{k}+ i (g-1) 
\mu \mathbf{V}_{k}\cdot\mathbf{S} $. \  $\mathbf{A}$ is the usual $U(1)$ field and $\mathbf{V}$ is the 
sought after tensor
field associated with the $SU(2)$ gauge or spin 
charge. \ This way of writing the covariant
derivative clearly displays that charge currents are associated with the
charge $e$ of the electrons while spin currents are associated with the
magnetic moment $\mu$. \ Therefore a 
spin current $\textsl{j}_{k}
=\textsl{j}_{k}^{\alpha} S_{\alpha}$ (summation implied) 
may exist 
in a system with zero charge but
nonzero magnetic moments and will satisfy a 'continuity' equation,
$\mathcal{D}_{k}\textsl{j}^{k} =0; \; k=0,1,2,3$. \ This is the result 
we show in this paper within a linear response approach. \ Hence 
in the following we neglect second order terms in the potential. This is 
equivalent to taking $e$ and $\mu$ as small parameters. 

\ The case of a constant electric field in the $xy$-plane 
and in the $x$-direction has
been well studied in the recent literature. \ Therefore in the 
following, we
focus on the homogeneous time-dependent case. \ The requirement 
of gauge
invariance under $U(1)\times SU(2)$ for the Rashba action
requires that   
$ \mathbf{V}_{0} = - \mathbf{B}$
and $
V_{k \alpha}=-\in_{k \alpha\beta}E_{\beta}$ for $k=x,y$ 
and $\alpha,\beta =1,2,3 $
(We use Greek or numerical indexes to denote spin 
components).  \ The Rashba coupling $\lambda$ can be
thought of as due to a fictitious electric field perpendicular 
to the $xy-
$plane of the 2DEG, $
\lambda=\frac{e}{4m^{2}}E_{z}$. \ The Rashba interaction 
is therefore replaced by $
H_{SOC}=\lambda \left(  p_{y}\sigma_{1}-p_{x}\sigma_{2}\right)  
-\frac{e}{4m^{2}
}p_{x}\sigma_{3}E_{x}$. \ For an electric field 
$\mathbf{E}= - \frac{1}{c} \partial\mathbf{A} /
\partial t$ in the $x$-direction, the interaction term 
becomes to first order in
the potential
\begin{equation}
H_{int} = \frac{e}{m} A_{x} (\mathbf{r},t) ( p_{x} - {\lambda m} \sigma_{2}).
\end{equation}
\ In the following we  
take $E_{z}>>E_{x}$ and hence the additional spin orbit coupling 
 due
to the E field in the $x$-direction can be neglected. \ The $E_z$ field 
will therefore be treated as a background field in the following. 

\ The generating functional of the theory is given by
\begin{eqnarray}
\lefteqn{Z\left[  \mathbf{A},\mathbf{V}\right]  =} \nonumber \\
&& \int D\psi^{\dagger}D\psi\exp\left\{
i\int dtd^{2}x\;\left[  \psi^{\dagger}\left(  t,\mathbf{r}\right)  \left(
G^{-1}\right)  \psi\left(  t,\mathbf{r}\right) \right. \right. \nonumber \\
&& \left.\left. +\psi^{\dagger}\left(
t,\mathbf{r}\right)  \psi\left(  t,\mathbf{r}\right)  K\psi^{\dagger}\left(
t,\mathbf{r}\right)  \psi\left(  t,\mathbf{r}\right)  \right]  \right\},
\end{eqnarray}
where we have included scattering due to impurities by a potential
$U$ 
, with $\left\langle U(r)U(r^{\prime
})\right\rangle =K\delta\left(  r-r^{\prime}\right)  $
where $K=1/(m\tau)$ and $\tau$ is the scattering.\ The propagator 
of the theory is given by
\begin{eqnarray}
G^{-1}  & =&
 i\partial_{t}-\frac{1}{2m}\pi^{2}-\frac{(g-1) \mu}{2m}\left[
\mathbf{S}\cdot\mathbf{V}_{k},\pi_{k}\right]  _{+} - g \mu \mathbf{S}\cdot
\mathbf{V}_0 ,
\end{eqnarray}
where  $\left[ , \right]_+$ is the anti-commutator 
operation,
$\pi_{l}=p_{l}+eA_{l}$ and $g=2$ for an electron in 
vacuum.  \ In the absence of impurity scattering, the path 
integral can be
easily integrated to get the effective action of the theory
\begin{equation}
\Gamma^{eff}\left[  \mathbf{A},\mathbf{V}\right]  =-i\;{\rm Tr}\;\log i\;G^{-1},
\end{equation}
where the trace is over space and spin indexes. \ 
 \  The case 
of nonzero-potential in the ladder approximation has 
been treated in
Ref.~[\onlinecite{mish}] using a Boltzmann-type 
approach.  \ To simplify our 
discussion,  we  set the impurity 
potential to zero in the rest of this paper since 
 our aim here 
is mainly to demonstrate the spin connection 
approach. \ We first briefly treat 
the case of 
a constant electric field and then discuss the 
inhomogeneous magnetic and electric 
fields. \ The spin current at $\mathbf{x}=(t,\mathbf{r})$ in the 
$y$-direction and polarized in the $z$-direction is
therefore given by \cite{IZ}
\begin{equation}
\frac{1}{\mu}\frac{\delta\Gamma^{eff}}{\delta V_{y 3}(\mathbf{x})}=-\textsl{j}
_{y3}(\mathbf{x}).
\end{equation}
This definition reduces to the usual definition of spin currents
 \cite{FS}.
\ In the linear response regime, we need 
only to calculate  $
G\left(  \mathbf{\ x},\mathbf{x}^{\prime}\right)  =G_{0}\left(
\mathbf{\ x},\mathbf{x}^{\prime}\right)  +\Sigma\left(  \mathbf{\ x}%
,\mathbf{x}^{\prime}\right)$  to first order
in $E_{x}$ where  $G_{0}$ is the $2\times2$ Green's function in the absence of the $E_{x}$
field, and its diagonal terms are given by  
$ G_{0}^{11}\left(  \omega\mathbf{\ },\mathbf{k}\right)
 =G_{0}^{22}\left(
\omega\mathbf{\ },\mathbf{k}\right) 
 =\left(\omega-\frac{k^{2}}{2m}\right)/\left[\left(  \omega-\frac{k^{2}}{2m}
 \pm i\epsilon\right)  ^{2}-\lambda^{2}k^{2}\right]$,
 while the off-diagonal ones are   $
G_{0}^{12}\left(  \omega\mathbf{\ },\mathbf{k}\right)  =G_{0}^{21\ast
}\left(  \omega\mathbf{\ },\mathbf{k}\right) 
 =\left( -i\lambda \left(  k_{x}-ik_{y}\right)\right)/\left[
\left(  \omega-\frac{k^{2}
}{2m} \pm i\epsilon\right)  ^{2}-\lambda^{2}k^{2}\right]$.
\ The poles of the Green function $G$ give the energy of the spin up and spin
down states of the system. \ Since $G_{11}^{0}=G_{22}^{0}$, only the $\Sigma
$-term contributes to the $(y 3)$-component of the spin current which
becomes
\begin{equation}
\textsl{j}_{y 3}=-\frac{1}{2 i m}{\rm Tr}\left\{  \sigma_{3}p_{y}G^{0}\left(
p_{x} + \lambda\;m\;\sigma_{2}\right)  \; K \;G^{0}\right\}
\end{equation}
with $
K = \frac{e}{2m}\left\{  A_{x},p_{x}\right\} $. \ To 
find the spin current, we go to Euclidean time $t_E = i t$.       \ 
The Fourier transform of the spin current for uniform fields is  
\begin{eqnarray}
\textsl{j}_{y 3}\left(  \omega\right)   & = & 
  =\frac{e\lambda}{4 \pi m} E ( \omega )\int_{k_{-}}^{k_{+}}dk\frac{k^{2}}{4\lambda^{2}%
k^{2}+\omega^{2}},
\end{eqnarray}
where $k_{\pm}=k_{F}\pm m\lambda$ are the momenta 
of the two bands and $k_F$ is the Fermi momentum, Fig. 
\ref{FermiSurface}.  \ Only states below the 
Fermi energy are occupied.\ The
static limit is therefore given by
\begin{equation}
\frac{\textsl{j}_{y 3}(0)}{E(0)}
=(g-1)\frac{\left|  e\right|  }{8  \pi \hbar },
\end{equation}
where we have restored physical units.
\ This result is equal to
  the one derived by Sinova {\em et al.}~\cite{sinova} and 
others \cite{loss} in the absence of any scattering.  \ In the presence 
of impurities, it was 
shown in Refs.~[\onlinecite{mish}] and [\onlinecite{chalaev}] that
  no matter how \textit{small} the 
scattering by the  impurities it 
cancels the spin Hall current. \ Hence it is natural to ask what 
happens if the electric or magnetic fields are not uniform 
as is usually the case 
in real devices. 

\ In order to shed some light on this question, we extend our treatment to non-uniform vector potential 
fields. \ We  only calculate spin currents in the $(y 3)$ 
direction. \ We  allow for both an electric field and a magnetic 
field to be applied to the Rashba electron gas. \ The vector potential 
$\mathbf{A}=\left( A_x , A_y ,A_z \right)$ is allowed to 
depend on both plane coordinates, $x$ and $y$. \ To calculate the 
spin current in the static limit, we use a gradient expansion to the 
effective action. \ In $QED_{2+1}$ such an expansion in momentum space 
gives 
rise to the well-known Chern-Simons term.\cite{das2} \ In our 
non-relativistic case, a real-space expansion is much easier
to carry out than in momentum space and this is 
what we do here. \ In the following, we keep only terms up to second order 
 derivatives in 
the vector potential.\ The 
calculation is carried out in the transverse gauge, 
$ \mathbf{k}\cdot \mathbf{A}(k)=0 $. \ 
\ The spin current in the $(y 3)$ direction  has two contributions, 
one coming from the kinetic term and the other from the Zeeman 
term. \ The kinetic term gives a contribution of the form
\begin{eqnarray}
J_{\mathbf{A}}\left(  \mathbf{r},\omega\right)  &=&  \left( g- 1 \right) \left(
a_{1}\left(  \omega\right)
A_{x}\left(  \mathbf{r},\omega\right)  +b_{1}\left(  \omega\right)
\frac{\partial^{2}A_{x}\left(  \mathbf{r},\omega\right)  }{\partial x^{2}%
} \right. \nonumber \\
&& \left. + b_{2}\left(  \omega\right)  \frac{\partial^{2}A_{x}\left(  \mathbf{r}%
,\omega\right)  }{\partial y^{2}}+c_{1}\left(  \omega\right)  \frac
{\partial^{2}A_{y}\left(  \mathbf{r},\omega\right)  }{\partial y\partial x} \right).
\end{eqnarray}
where the frequency dependent coefficients are given by
\begin{equation}
a_{1}\left(  \omega\right)    =\frac{i e\lambda}{ 2 m}
\int\frac{d^{2}k_{1}}%
{(2\pi)^{2}}\frac{d\omega_{1}}{2\pi}tr\left\{  \sigma^{3}G^{0}\left(
\mathbf{k}_{1},\omega_{1}\right)  \sigma^{2}G^{0}\left(  \mathbf{k}_{1}%
,\omega_{1}-\omega\right)  \right\}  k_{1}^{y},
\end{equation}
\begin{eqnarray}
b_{1}\left(  \omega\right)   & =&  \frac{i e}{4 m}
\int\frac{d^{2}k_{1}}{(2\pi
)^{2}}\frac{d\omega_{1}}{2\pi}tr\left\{  \sigma^{3}G^{0}\left(  \mathbf{k}%
_{1},\omega_{1}\right)  \partial_{k_{x}}^{2}G^{0}\left(  \mathbf{k}_{1}%
,\omega_{1}-\omega\right)  \right\}  k_{1}^{x}k_{1}^{y} \\
&& -\frac{ie\lambda}{4 m}\int\frac{d^{2}k_{1}}{(2\pi)^{2}}\frac{d\omega_{1}}%
{2\pi}tr\left\{  \sigma^{3}G^{0}\left(  \mathbf{k}_{1},\omega_{1}\right)
\sigma^{2}\partial_{k_{x}}^{2}G^{0}\left(  \mathbf{k}_{1},\omega_{1}%
-\omega\right)  \right\}  k_{1}^{y},\nonumber
\end{eqnarray}
\begin{eqnarray}
b_{2}\left(  \omega\right)     &=& \frac{i e}{4 m}
 \int\frac{d^{2}k_{1}}{(2\pi
)^{2}}\frac{d\omega_{1}}{2\pi}tr\left\{  \sigma^{3}G^{0}\left(  \mathbf{k}%
_{1},\omega_{1}\right)  \partial_{k_{y}}^{2}G^{0}\left(  \mathbf{k}_{1}%
,\omega_{1}-\omega\right)  \right\}  k_{1}^{x}k_{1}^{y}\\
&& -\frac{ie\lambda}{4 m}\int\frac{d^{2}k_{1}}{(2\pi)^{2}}\frac{d\omega_{1}}%
{2\pi}tr\left\{  \sigma^{3}G^{0}\left(  \mathbf{k}_{1},\omega_{1}\right)
\sigma^{2}\partial_{k_{y}}^{2}G^{0}\left(  \mathbf{k}_{1},\omega_{1}%
-\omega\right)  \right\}  k_{1}^{y},\nonumber 
\end{eqnarray}
\begin{eqnarray}
c_{1}\left(  \omega\right)     &=& \frac{i e}{2 m}
 \int\frac{d^{2}k_{1}}{(2\pi)^{2}%
}\frac{d\omega_{1}}{2\pi}tr\left\{  \sigma^{3}G^{0}\left(  \mathbf{k}%
_{1},\omega_{1}\right)  \partial_{k_{x}}\partial_{k_{y}}G^{0}\left(
\mathbf{k}_{1},\omega_{1}-\omega\right)  \right\}  k_{1}^{y2}\\
&& +\frac{ie\lambda}{2 m}\int\frac{d^{2}k_{1}}{(2\pi)^{2}}\frac{d\omega_{1}}{2\pi
}tr\left\{  \sigma^{3}G^{0}\left(  \mathbf{k}_{1},\omega_{1}\right)
\sigma^{1}\partial_{k_{x}}\partial_{k_{y}}G^{0}\left(  \mathbf{k}_{1}%
,\omega_{1}-\omega\right)  \right\}  k_{1}^{y}. \nonumber
\end{eqnarray}
\ The first term, $a_1(\omega)$, is the original electric 
field related
term found above for 
the uniform electric field and is given (in the imaginary-time) by
\begin{eqnarray}
a_1 (\omega) &= &\frac{ie \omega}{32 \pi m\lambda^2}\left( 4m\lambda^2\right. \\
&& \left. + \omega \tan^{-1}
\left[ \frac{2\lambda(k_F - m\lambda)}{\omega}\right] - \omega \tan^{-1}
\left[\frac{2\lambda(k_F + m\lambda)}{\omega}\right] \right).\nonumber
\end{eqnarray}
\ Hence in the static limit, $\omega \rightarrow 0$, 
this term vanishes and hence we choose not to include the electric 
field contribution below which is the same as in the uniform 
case. \ In the non-static case time dependent derivatives 
terms of the electric field also appears in the time-dependent
 spin current.\ The 
remaining non-zero $\omega$-dependent  terms are too long to be given here.
\ For the $(y 3)$ direction, the Zeeman contribution 
is nonzero only for a magnetic field with a normal component
 to the 
plane,
\begin{eqnarray}
\lefteqn{J_{{Zeeman}}\left(  \mathbf{r},\omega\right)  = 
-\frac{g e}{8 m^{2}
}} \\
&& \times\partial_{y}B_{z}\left(  \mathbf{r},\omega\right)  \int\frac{d^{2}k_{1}
}{(2\pi)^{2}}\frac{d\omega_{1}}{2\pi}{\rm Tr}\left\{  \sigma^{3}G^{0}\left(
\mathbf{k}_{1},\omega_{1}\right)  \sigma^{3}\partial_{k_{y}}G^{0}\left(
\mathbf{k}_{1},\omega_{1}-\omega\right)  \right\}  k_{1}^{y}. \nonumber
\end{eqnarray}
\ Summing all these terms and making use of the transverse condition
for the gauge field $\mathbf{A}$, we get in the static limit the following
simple result
\begin{equation}
j_{2}^{3}\left(  \mathbf{r},0\right)  = \frac{\left( 2 g - 1 \right) \mu}{16 \pi } \;
\partial_{y}B_{z}\left(  \mathbf{r},\omega=0\right).
\end{equation}
\ This is the main result of this communication which is 
really the Chern-Simons equivalent of the Rashba model and is 
gauge-invariant. \ The $-1$ in $(2g-1)$ can be traced back to  
the Thomas correction. \ The y-partial derivative 
can be understood by going to the rest frame of the 
{\it spin}, i.e., zero torque frame, where the SO is 
seen to give rise to a force in the y-direction which 
is the source of the spin current in a direction perpendicular 
to the charge current. \ We observe that in 
non-uniform magnetic fields, the conductivity is 
proportional to the magnetic moment
of the conduction electrons and is
 independent of the Rashba 
coupling as was the case for uniform electric fields. \ In this 
reformulation of the Rashba model this fact is easily understood 
in terms of $SU(2)$ gauge invariance. \ Using Maxwell's equations, 
it can be shown that the spin conductivity is approximately 
equal to the charge conductivity multiplied by the magnetic 
moment.   \ Moreover, it should 
be stressed that the constant that appear in front of 
the magnetic field gradient is really a property of the 
topology of the two-dimensional $k-$space.

\ The fact that a non-uniform magnetic field gives rise 
to a  spin current is not surprising. \ Loss and 
Goldbart \cite{Loss} 
were able to show that in a ring with a non-uniform Zeeman 
term can give rise
to an equilibrium  , or persistent,
 charge and spin currents.\ In their system, however, the 
persistent current is due to the phase
coherence of the wavefunction. In the presence 
of a non-uniform magnetic field, a Berry's phase is 
added to the usual Aharonov-Bohm phase. \ Enforcing 
single-valuedness of the wavefunction results in 
persistent charge and spin currents. \ In contrast, there is here 
no real-space topological 
constraint as is the case for a mesoscopic ring.  \  Similarly 
 in Ref.~[\onlinecite{rebei}], it was  shown 
that a persistent spin current is possible under the action 
of an s-d exchange term induced by a non-uniform 
magnetization. \ The magnetization in that  case
plays similar role to that of the non-uniform $B$-field in 
the Zeeman term. \ In Fig. 1, the $k_x = k_y = 0$ point is 
a degeneracy point and hence will give rise to a point 
source term for the Berry curvature, Eq.(9) in Ref. 
\onlinecite{haldane}. \ Hence topologically we are 
dealing with a ring structure in k space as opposed to real space 
in the above two  examples.

\ In summary, we have solved for the spin 
Hall conductivity using a path
integral approach. The spin current was defined 
via a spin $SU(2)$ 
gauge potential. \ Our main two results
are first of all that the conductivity in 
the static limit for a 
uniform electric field was found to be the same as the 
one calculated by the Kubo formula. \ Second, we found 
that the inclusion of   
 an inhomogeneous magnetic 
field in the $z-$direction  alters this result by adding a 
new component to the  spin current. \ This component
is proportional to the classical force
on a magnetic dipole in a non-uniform magnetic field
 and is still independent 
of the Rashba coupling. \ Hence impurity scatterings are unlikely 
to suppress spin currents generated by non-uniform fields.

\ {\it Note added - \;} Recently, it has come to our 
attention that the
 idea put forward here
of using the $U(1)
\times SU(2)$ symmetry as a basis for
 describing spin currents has been also noted
by Schmeltzer \cite{dave}. \ The reader is refrerred to his work
for a more detailed discussion of the spin current equation.

\bigskip

 We appreciate initial discussions with G. W. Bauer and 
E. Rossi. \ We thank E. Simanek for useful 
comments. \ We also acknowledge 
useful discussions with A. MacDonald, Q. Niu, P. Goldbart, and J. Hohlfeld.


\begin{thebibliography}{99}


\bibitem{das}I. Zutic, J. Fabian, and S. Das Sarma, Rev. Mod. Phys.
\textbf{76}, 323 (2004).

\bibitem{rashba1}E. I. Rashba, Phys. Rev. B \textbf{68}, 241315(R) (2003).

\bibitem{rashba2}E. I. Rashba, cond-mat/0404723.

\bibitem{FS}J. Frohlich and U. M. Studer,
Rev. Mod. Phys. \textbf{65}, 733 (1993).

\bibitem{sinova}J. Sinova, D. Culcer, Q. Niu, N. A. Sinitsyn, T. Jungwirth,
and A. H. MacDonald, Phys. Rev. Lett. \textbf{92}, 126603 (2004).

\bibitem{mish}E. G. Mishchenko, A. V. Shytov, and B. I. Halperin, Phys. Rev.
Lett. \textbf{93}, 226602 (2004).

\bibitem{inoue}J. I. Inoue, G. E. W. Bauer, and L. W. Molenkamp, Phys. Rev.
\textbf{B} 70, 041303(R) (2004).

\bibitem{raimondi}R. Raimondi and P. Schwab, Phys. Rev. B \textbf{71}, 033311 (2005).

\bibitem{loss}S. I. Erlingsson, J. Schliemann, and D. Loss, Phys. Rev.B
\textbf{71}, 035319 (2005).

\bibitem{chalaev}O. Chalaev and D. Loss, Phys. Rev. \textbf{B} 71, 245318.

\bibitem{bernevig}B. A. Bernevig, Phys. Rev. \textbf{B} 71, 073201 (2005).

\bibitem{chao}A. G. Mal'shukov and K. A. Chao, Phys. Rev. \textbf{B} 71
121308(R) (2005).

\bibitem{chao2} A. G. Mal'shukov, C. S. Tang, C. S. Chu, and K. A. Chao, Phys. 
Rev. \textbf{B} 68, 233307 (2003); C. S. Tang, A. G. Mal'shukov, and 
K. A. Chao, Phys. Rev. B \textbf{71}, 195314 (2005).

\bibitem{niu}P. Zhang, J. Shi, D. Xiao, and Q. Niu, Phys. Rev. Lett. 
\textbf{96}, 076604 (2006); cond-mat/0503505v1.


\bibitem{sun} Q-F. Sun and X. C. Xie, Phys. Rev. \textbf{72}, 245305 
(2005).

\bibitem{schwinger}J. Schwinger, Phys. Rev.\textbf{82}, 664 (1951).

\bibitem{IZ}C. Itzykson and J-B. Zuber, Quantum Field Theory, McGraw-Hill ,
New York (1980).

\bibitem{bychkov}Y. A. Bychkov and E. I. Rashba, J. Phys. C \textbf{17}, 6039 (1984).

\bibitem{das2} K. S. Babu, A. Das, and P. Panigrahi, Phys. Rev. D \textbf{36}, 
3725 (1987).
\bibitem{Loss} D. Loss and P. Goldbart, Phys. Rev. B \textbf{45}, 13544
  (1992).

\bibitem{rebei} A. Rebei, W.N.G. Hitchon, and G. J. Parker, Phys. Rev. B 
\textbf{72}, 064408 (2005).

\bibitem{haldane} F. D. M. Haldane, Phys. Rev. Lett. \textbf{93}, 
206602 (2004).

\bibitem{dave} D. Schmeltzer, cond-mat/0509607v2.

\end{thebibliography}
\end{document}